# Reinforcement Learning-Based Adaptive Insulin Advisor for Individuals with Type 1 Diabetes Patients under Multiple Daily Injections Therapy*

Qingnan Sun, *Student Member, IEEE*, Marko V. Jankovic, Stavroula G. Mougiakakou, *Member, IEEE*

*Abstract*— The existing adaptive basal-bolus advisor (ABBA) was further developed to benefit patients under insulin therapy with multiple daily injections (MDI). Three different *in silico* experiments were conducted with the DMMS.R simulator to validate the approach of combined use of self-monitoring of blood glucose (SMBG) and insulin injection devices, e.g. insulin pen, as are used by the majority of type 1 diabetes patients under insulin therapy. The proposed approach outperforms the conventional method, as it increases the time spent within the target range and simultaneously reduces the risks of hyperglycaemic and hypoglycaemic events.

## I. INTRODUCTION

Type 1 diabetes (T1D) is caused by the destruction of pancreatic beta cells and is a metabolic disease characterised by high levels of blood glucose (hyperglycaemia). If they are to maintain their glucose levels within a healthy range (e.g. 70 md/dl to 180 mg/dl), type 1 diabetic patients need external insulin delivery to compensate for the increase in blood glucose, especially due to the intake of carbohydrates (CHOs).

There are three important and essential steps in achieving appropriate regulation of blood glucose, i.e. a) monitoring blood glucose, b) calculation of the required amount of insulin, and c) delivery of the insulin dose. With developments in technology, more and more approaches for glucose monitoring have become available, e.g. SMBG (also known as the blood glucose meter, BGM), continuous glucose monitoring (CGM) systems, and flash glucose monitoring (FGM) [1], [2]. SMBG is the conventional and most widely used glucose monitoring device, and measures the blood glucose level with one drop of finger blood. According to the NICE guideline [3], type 1 diabetic adults are recommended to test glucose levels at least four times a day, i.e. before each meal and before going to bed. CGM devices can provide glucose measurements every few minutes, by using a subcutaneous miniaturised sensor and wireless communication between the sensor and the mobile devices. Recently, FGM device is a hybrid between meters and CGMs, and has attracted recent attention. As described in [1], FGM uses a sensor implanted in the arm that a user (health care provider or patient) scans with a specialised reader to record glucose levels, trends, and patterns. In contrast to CGM, FGM only provides a trend graph if it has been swiped in the preceding eight hours, and there is no possibility of generating real time glycaemia alarms [4].

The insulin pen and the continuous subcutaneous insulin infusion (CSII) pump deliver insulin to diabetic patients. The corresponding insulin therapies are designated as "multiple daily injections (MDI) therapy" and "insulin pump therapy", respectively. Total insulin normally consists of basal insulin and bolus insulin [5]. The former maintains blood glucose during the fasting period, while the latter compensates for the increase in blood glucose caused by CHO intake. For the conventional insulin pen user, long-acting insulin is used for basal insulin, and short or rapid insulin for bolus insulin. With the CSII pump, only rapid insulin is used. However, depending on the infusion rate, the infused insulin can be designated as either the "basal rate" (BR, very low infusion rate during the whole day) or bolus dose (fast infusion rate, before meals) [6].

The patient-specific values of basal insulin and CIR (CHO-to-insulin ratio, calculation of the size of the bolus) are initially given by the healthcare specialist. Then, the values can be updated by different approaches, based on the blood glucose concentration of the patients. Zisser *et al.* [7] demonstrated clinically that run-to-run (R2R) control can be used to manage meal-related insulin in subjects with T1D. In [8], the authors introduced an R2R algorithm, which used only post-prandial blood glucose measurements to adjust bolus insulin. The system remained stable, but with large uncertainty. Then, the R2R algorithm was applied to use 5 glucose measurements as input to successfully adjust basal insulin [9]. Some other studies used CGM measurements as the inputs for control algorithms. For instance, Herrero *et al.* introduced a novel method based on case-based reasoning and an R2R algorithm for automatic adjustment of the bolus calculator [10], [11] and for adaptation of basal insulin [12]. Toffanin *et al.* introduced an R2R algorithm for basal insulin adaptation, and this exhibited good performance and stability on a population of 100 virtual diabetic adults [13]. In [14] and [15], the authors proposed adjusting the bolus and basal insulin simultaneously.

In the aforementioned studies, either CGM or a CSII pump, or both were required. In comparison to SMBG, the CGM device provides many more glucose measurements per day to

*Research supported in part by the Swiss Commission of Technology and Innovation (CTI) under Grant 18172.1 PFLS-LS.

Qingnan Sun is with the ARTORG Center for Biomedical Engineering Research, University of Bern. Murtenstrasse 50, CH-3008 Bern, Switzerland (phone: +41-31-632-7596; e-mail: qingnan.sun@artorg.unibe.ch).

Marko V. Jankovic is with the Department of Emergency Medicine, Bern University Hospital "Inselspital", Bern, Switzerland and with the ARTORG Center for Biomedical Engineering Research, University of Bern. Murtenstrasse 50, CH-3008 Bern, Switzerland (e-mail: Marko.jankovic@artorg.unibe.ch).

Stavroula G. Mougiakakou is with the ARTORG Center for Biomedical Engineering Research, University of Bern, Murtenstrasse 50, CH-3008 Bern, Switzerland and the Department of Diabetes, Endocrinology, Clinical Nutrition and Metabolism, Bern University Hospital "Inselspital", Bern, Switzerland (phone: +41-31-632-7592; e-mail: stavroula.mougiakakou@artorg.unibe.ch).

monitor the glucose concentration. The CSII pump enables easy adjustment of the insulin infusion rate during the day, which is not possible by using an insulin pen. This advantage is especially meaningful for basal insulin infusion, since different basal rate profiles can be implemented with the CSI pump. However, since the majority of diabetic patients use SMBG or an insulin pen [16]–[18], algorithms for insulin adjustment should also be investigated for these approaches.

In the previous study [19], we introduced an algorithm based on reinforcement learning for the automatic adjustment of BR and CIR, in the context of the artificial pancreas (AP), which used a combination of CGM and the CSII pump. In the continuation of this study [20], the adaptive basal-bolus algorithm (ABBA) was optimised in various respects, including the capability to use four SMBG measurements per day as system inputs. As a further extension, in the present study, we investigate whether it is possible to apply the algorithm with the combination of SMBG device and insulin pen. The current preliminary *in silico* work was conducted on T1D patients, but may also be used for type 2 diabetes (T2D) patients.

## II. METHODOLOGY

In this section, we present the method of the adaptive basal-bolus advisor based on reinforcement learning, which works with SMBG devices and the insulin pen. The baseline bolus advisor is also introduced.

### A. Bolus Advisor

The Bolus Advisor (BA) uses a simple algorithm to support the diabetic patient in deciding the amount of meal-related insulin, i.e. bolus insulin. Usually, the bolus amount provided by BA contains two parts: the meal insulin part and the correction insulin part [21]. The former is directly calculated from the estimated CHO amount and the CIR value, while the latter is an adjustment of the meal insulin part based on the current blood glucose level. The algorithm of a BA can be described as:

$$\text{Bolus insulin} = \text{Meal insulin} + \text{Correction insulin}, \quad (1)$$

$$\text{Bolus insulin} = \text{CHO/CIR} + (\text{BG} - \text{Target})/\text{CF}, \quad (2)$$

where BG is the current blood glucose value, Target is the target glucose value, and CF is the correction factor. Both CIR and CF are subject-specific metabolic parameters, which are given by the healthcare specialist on the basis of their estimation.

### B. ABBA for SMBG and MDI therapy

In our previous study [20], a dual model adaptive basal-bolus advisor (ABBA) was introduced and the mathematical details were described. ABBA can use either CGM or SMBG measurements as system input. The output of the system, i.e. basal rate and bolus insulin, was delivered by a CSII pump. In this case, the basal insulin was slowly infused as a flat profile during the whole day, and the bolus insulin was infused in the course of 5 minutes.

In the case of MDI therapy, e.g. with the insulin pen, two main differences were considered for modifying the algorithm, a) instead of a infusion rate, basal insulin needed to be injected once per day, b) both basal insulin and bolus insulin were injected within a short time.

Thus the following calculation was applied to convert the basal rate, as calculated with ABBA, into the proper amount of long acting insulin for a single injection:

$$\text{Basal insulin} = \text{basal rate} * 1440, \quad (3)$$

where 1440 is the number of minutes in one day. On the other hand, it was no longer necessary to divide the total bolus amount by the infusion duration, e.g. 5 minutes, since the whole bolus insulin needs to be injected together. In this study, the injection time for both basal insulin (long acting insulin) and bolus insulin (rapid insulin) was one minute.

The Transfer Entropy (TE) method for initialisation [22] was not applied, since pump therapy was required for the latter method. Thus, the control policy was initialised to 0.5, which is the mean value of its range (0.1). From the point of view of the algorithm, the rest of ABBA remained the same as described in [20].

## III. EXPERIMENTAL PROTOCOL

This section describes the experimental protocol, which means the simulation environment, experiment scenarios, meal protocol, etc..

### A. Simulation Environment

The Diabetes Mellitus Metabolic Simulator for Research (DMMS.R) was used to evaluate the aforementioned algorithm. The DMMS.R simulator is a computer application designed for conducting clinical studies in virtual subjects [23]. In contrast to the Type 1 Diabetes Metabolic Simulator (T1DMS) [24], the newly released DMMS.R simulator provides more cohorts of *in silico* subject population (i.e. T1D, T2D or Pre-Diabetes). Furthermore, besides rapid-acting insulin, DMMS.R also supports simulation with long-acting insulin or oral medications, which introduces more possibilities by using different treatments for the *in silico* experiments. To the best of our knowledge, this is the first published study of *in silico* evaluation to be conducted with the DMMS.R simulator for diabetic patients with the SMBG device and MDI therapy.

In this study, the 10 virtual adult subjects of T1D cohort provided by the DMMS.R simulator were used for simulation. The blood glucose levels were measured by the virtual SMBG device. Long-acting insulin and rapid-acting insulin were used to reflect insulin treatment for insulin pen users.

### B. Experimental scenarios

The total duration of the *in silico* experiments was 15 days. The first day was excluded from the evaluation, since the insulin on board was zero at the very beginning of each experiment, which could lead to more hyperglycaemia on the first day. The second day (D2) to the eighth day (D8), in total seven days (one week), was defined as W1 and the treatment used was suggested by BA. In the meantime, ABBA updated the features based on the daily glucose measurements. From the ninth day (D9) to the fifteenth day (D15), i.e. in week 2 (W2), ABBA provided daily suggestions for the long-acting insulin amount, as well as the bolus amount for the main meals.

TABLE I. GLUCOSE LEVELS (MEAN ± STANDARD DEVIATION)

|  | E1 | | E2 | | E3 | |
| --- | --- | --- | --- | --- | --- | --- |
|  | *BA* | *ABBA* | *BA* | *ABBA* | *BA* | *ABBA* |
| % in Target | 73.5±8.6 | 83.9±9.3 | 76.8±6.0 | 91.7±5.2 | 76.9±6.0 | 91.7±5.3 |
| % in Hypo | 0.8±1.0 | 0.7±1.1 | 8.9±4.3 | 1.8±0.9 | 9.1±4.3 | 1.8±0.8 |
| % in Severe Hypo | 0.1±0.4 | 0.0±0.1 | 4.5±4.0 | 0.7±0.9 | 3.9±3.6 | 0.6±1.0 |
| % in Hyper | 25.6±8.8 | 15.4±8.6 | 9.8±4.7 | 5.9±4.7 | 10.1±4.8 | 5.9±4.7 |
| % in Severe Hyper | 0.0±0.0 | 0.0±0.0 | 0.0±0.0 | 0.0±0.0 | 0.0±0.0 | 0.0±0.0 |

The DMMS.R simulator provides for virtual T1D subjects with the patient-specific "optimal" basal insulin and CIR, which can bring the virtual subjects to a relatively well controlled condition. To reflect the bias that the healthcare specialist may have when estimating "optimal" values, an uncertainty of 10% of the given value was applied. Three experiments were designed to evaluate the performance of the algorithms from different aspects:

- E1: the amount of "optimal" basal and bolus insulin was decreased by 10%;
- E2: the amount of both "optimal" basal and bolus insulin was increased by 10%;
- E3: the amount of both "optimal" basal and bolus insulin was considered to have a uniformly distributed uncertainty of ±10%.

The experiments were conducted with the virtual adult T1D cohort. There were in total 11 virtual subjects, including 10 individual subjects as well as one average subject. Subject number 7 was aged 47 with a body weight of only 46 kg and exhibited strikingly high fluctuations in blood glucose in comparison with the other subjects. A 10% increase on top of the "optimal" treatment caused an extreme reaction during BA the phase, so this subject was excluded from the scenarios.

### C. Meals, Insulin Sensitivity, Glucose Measurements and Insulin Delivery

Four meals per day, presented as breakfast, lunch, dinner and bed time snack, were considered in the *in silico* evaluation. The meal timings and the CHO amounts are: 07:00 h (50 g), 12:00 h (80 g), 18:30 h (70 g) and 23:00 h (15 g). Furthermore, in order to reflect the error in the estimation of the patients' CHO, a uniformly distributed uncertainty of ±50% was considered for the CHO amount used in calculating the bolus dose.

The intraday variability of insulin sensitivity was considered in the sense of the "dawn phenomenon". This refers to periodic episodes of hyperglycaemia occurring in the early morning hours before and after breakfast [25]. As implemented in [20], in the present study, SI also dropped every day between 04:00 and 08:00 to 50% of its original value, and SI ramped up or down within around 30 minutes.

Four glucose measurements per day, which were measured with virtual SMBG devices, were required. Three of them were pre-meal measurements and were measured 20 minutes before the main meals. The bedtime measurement took place at 23:00h.

Long-acting insulin, which was used to keep blood glucose under control throughout its daily routine, was injected once per day at 23:00 h, i.e. directly after the last measurement of the day. The rapid-acting insulin, i.e. the bolus dose, was injected after the pre-meal measurements, 20 minutes before the main meals. This configuration reflected the general injection habits of users of insulins.

### D. Evaluation metrics

The evaluation of the performance of the experimental scenarios was based on the analyses of the blood glucose level of the virtual patients. Different metrics were implemented to assess the performance. The most widely used parameter is the percentage of time in different blood glucose ranges: percentage time in glucose target range [70-180] mg/dl; percentage time in hypoglycaemia [50-70) mg/dl; percentage time in severe hypoglycaemia <50 mg/dl; percentage time in hyperglycaemia (180-300] mg/dl; and percentage time in severe hyperglycaemia >300 mg/dl.

Furthermore, two glycaemic indices, LBGI (Low Blood Glycaemic Index) and HBGI (High Blood Glycaemic Index) were considered. LBGI indicates the risk of hypoglycaemia, and HBGI indicates the risk of hyperglycaemia.

## IV. RESULTS AND DISCUSSION

Table I presents the results of the three scenarios of the *in silico* experiments. During the BA phase, since in E1 less initial insulin was given, the subjects suffered much hyperglycaemia. In contrast, more initial insulin in E2 reduced the percentage in hyperglycaemia but also increased the time spent in hypoglycaemia. In comparison with BA, and in each scenario, ABBA reduced the percentage of time in the (severe) hypo- and hyperglycaemic ranges and increased 10% to 15% in the target range.

Fig. 1 shows the daily LBGI and HBGI values of the experiments E1 to E3. In the case of E1, in the second week (W2), HBGI was reduced to under 5 (minimal risk), while LBGI remained within 1 (minimal risk.). In both E2 and E3, HBGI remained at minimal risk during the whole process, while LBGI was reduced by ABBA from medium risk (2.5 to 5.0) to minimal risk (smaller than 1).

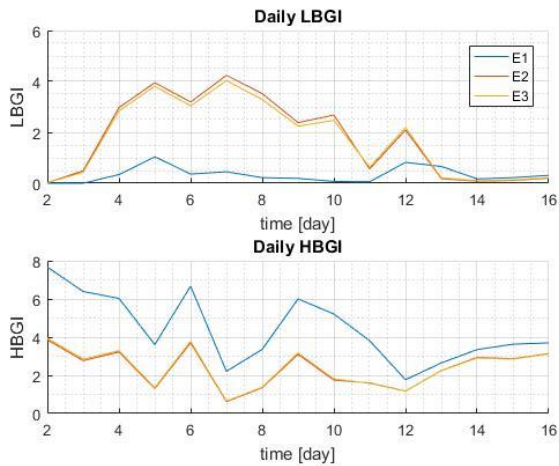

Figure 1. Daily LBGI and HBGI

## V. Conclusion

This paper presents the application of ABBA, an adaptive basal-bolus algorithm, to T1D patients who use the SMBG device and insulin pen. Three different experimental scenarios were designed to evaluation the algorithm in comparison with the conventional bolus advisor. The preliminary results of *in silico* trials with the DMMS.R simulator indicate that the algorithm is promising in the proposed applications. More experiments are needed to validate the algorithm with more variabilities and uncertainties, and the algorithm needs to be further optimised for extreme diabetic individuals. Furthermore, this approach has the potential to be applied for T2D patients who use SMBG and MDI therapy.